# Particle Image Velocimetry (PIV) Uncertainty Quantification Using Moment of Correlation (MC) Plane

Sayantan Bhattacharya[1], John J. Charonko[2], Pavlos P. Vlachos[1]

[1] School of Mechanical Engineering, Purdue University, USA.
[2] Physics Division, Los Alamos National Laboratory, USA.

# Abstract

We present a new uncertainty estimation method for Particle Image Velocimetry (PIV), that uses the correlation plane as a model for the probability density function (PDF) of displacements and calculates the second order moment of the correlation (MC). The cross-correlation between particle image patterns is the summation of all particle matches convolved with the apparent particle image diameter. MC uses this property to estimate the PIV uncertainty from the shape of the cross-correlation plane. In this new approach, the Generalized Cross-Correlation (GCC) plane corresponding to a PIV measurement is obtained by removing the particle diameter contribution. The GCC primary peak represents a discretization of the displacement PDF, from which the standard uncertainty is obtained by convolving the GCC plane with a Gaussian function. Then a Gaussian least-squares-fit is applied to the peak region, accounting for the stretching and rotation of the peak, due to the local velocity gradients and the effect of the convolved Gaussian. The MC method was tested with simulated image sets and the predicted uncertainties show good sensitivity to the error sources and agreement with the expected RMS error. Subsequently, the method was demonstrated in three PIV challenge cases and two experimental datasets and was compared with the published image matching (IM) and correlation statistics (CS) techniques. Results show that the MC method has a better response to spatial variation in RMS error and the predicted uncertainty is in good agreement with the expected standard uncertainty. The uncertainty prediction was also explored as a function PIV interrogation window size, and the MC method outperforms the other uncertainty methods.

# Nomenclature

$\sigma$ : Standard deviation

$\sigma_u$ : Standard uncertainty

$<R>$ : Ensemble averaged cross correlation plane

$\mathcal{F}$ : Forward Fourier transform

$R^*$: Spectral cross-correlation

$P_I$ : Particle image shape information

$G(x)$: Generalized Cross correlation (GCC)

$p(x)$: PDF of displacement.

$I_{XX}$: Second order moment about *x*- axis

$I_{YY}$: Second order moment about *y*-axis

$R_{conv}$: Gaussian convolved PDF plane

$N_{eff}$: Effective number of pixels contributing to correlation

$e_{Prana}$: Error in velocity measurements obtained using Prana processing

$e_{DaVis}$: Error in velocity measurements obtained using DaVis processing

$\sigma_{MC}^{x}$: Standard *x* uncertainty estimate using MC method

$\sigma_{MC}^{y}$: Standard *y* uncertainty estimate using MC method

$\sigma_{IM}$: Standard uncertainty estimate using IM method

$\sigma_{CS}$: Standard uncertainty estimate using CS method

# 1  Introduction

Particle Image Velocimetry (PIV) is a non-invasive quantitative fluid velocity measurement technique in which tracer particles are illuminated by a laser sheet, imaged by a high-speed camera, and the displacement of the particle patterns within an image sequence is estimated to resolve the velocity field. An overview of the development of PIV over the past 20 years is given by Adrian[1], and a comprehensive history can be traced in recent publications [2], [3]. Currently, the term PIV is used to encompass the extensive family of methods that are based on evaluating the particle patterns displacement using statistical cross-correlation of consecutive images with high number density flow tracers[2].

However, despite detailed investigation of potential error sources, the development of PIV methods did not involve simultaneous rigorous quantification of uncertainty for a given measurement. As a result there is currently no widely accepted framework for reliable quantification of PIV measurement uncertainty. The situation is exacerbated by the fact that PIV measurements involve instrument and algorithm chains with coupled uncertainty sources, rendering quantification of uncertainty far more complex than most measurement techniques. Also, knowing the uncertainty bound on each PIV vector is crucial in comparing experimental results with numerical simulations. Therefore, developing a fundamental methodology for quantifying the uncertainty for PIV is an important and outstanding challenge.

Recent developments in this field have led to several uncertainty estimation methods which can be broadly classified into indirect and direct uncertainty estimation algorithms.

## 1.1 Indirect methods

The indirect methods use pre-calculated calibration information to predict the measurement uncertainty. In the first such method published, Timmins et al. constructed an "Uncertainty Surface"(US) by mapping the effects of selected primary error sources such as shear, displacement, seeding density, and particle diameter to the distribution of the true errors for a given measurement[4]. This approach is analogous to a traditional instrument calibration procedure for standard experimental instruments. Ultimately, in order to comprehensively quantify the uncertainty, all possible combinations of displacements, shears, rotations, particle diameters, and other parameters must be exhaustively tested which can make this method computationally expensive. Moreover, many of the relevant parameters may not be easily obtained from a real experiment.

Charonko and Vlachos proposed an uncertainty quantification method based on the ratio of the primary peak height to the second largest peak (PPR) [5] in the correlation plane. Using this method, the uncertainty of PIV measurement can be predicted without *a priori* knowledge of image quality and local flow conditions. Reliable uncertainty estimation results using a phase-filtered correlation (RPC)[6] were shown, however for standard cross-correlation (SCC) techniques the uncertainty estimates were not as good. Also, the approach depends, like the uncertainty surface method, on calibration of the peak ratio to the expected uncertainty. Xue et al.[7] used an analogous approach to calibrate the measurement uncertainty with various other correlation plane signal to noise ratio (SNR) metrics (namely PRMSR, PCE and ENTROPY). The uncertainty coverage, which denotes the probability of measurement errors falling between the uncertainty bounds is used as a metric to compare the different uncertainty predictions. The SNR based uncertainty methods developed by Xue et al. showed an improved uncertainty coverage for both RPC and SCC. In another effort, the effective information contributing to the cross correlation plane primary peak was named the "Mutual Information (MI)"[8] and used to predict the PIV measurement uncertainty. The MI between a correlated image pair is an estimate of the effective number of correlating particles and thus higher MI should correspond to a lower uncertainty on the measured velocity. Xue et al. successfully used MI as an indirect metric to predict the uncertainty in a PIV measurement.

## 1.2 Direct methods

The uncertainty in a measurement can also be extracted directly from the image plane using the estimated displacement as a prior information. Sciacchitano et al. proposed a method to quantify the uncertainty of PIV measurement based on particle image matching (IM) or particle disparity [9]. The uncertainty of measured displacement is calculated from the ensemble of disparity vectors, which are due to incomplete matching between particle pairs within the interrogation window. This method accounts for random and systematic error; however peak-locking errors and truncation errors cannot be detected. In addition, the disparity can be calculated only for particles that are paired within the interrogation window, thus this method cannot account for the effects of in-plane and out-of-plane loss of particles. Finally, particle image pair detection can introduce additional sources of error and the method can be computationally expensive for high resolution images with higher seeding density.

Wieneke in his "Correlation Statistics"(CS) method computed the measurement uncertainty by relating the asymmetry in the correlation peak to the covariance matrix of intensity difference between two almost matching interrogation windows [10]. This is a more generalized image matching technique where the random error is estimated by the variance of pixel wise intensity difference and linked to the correlation function shape using the uncertainty propagation for a 3-point Gaussian fit. Due to pixel-wise matching, any loss of correlation due to out of plane motion or other possible error sources are taken into account. However the method is limited statistically in case of smaller window size and bigger particle image size.

In a comparative assessment of the methods, Sciacchitano et al. [11] compared these four methods for an experimental jet case. Four different cases were tested, each one having a dominant primary error source (shear, out-of-plane motion, particle size and seeding density). The authors established that for zero bias the RMS of the error distribution should match the RMS of the predicted uncertainty distributions and this was used as the basis of comparison. The results indicated a better uncertainty prediction and sensitivity to RMS error variation for the direct methods (CS and IM) in all four cases. Both the calibration-based methods underperformed. The PPR method showed less sensitivity, especially in the shear region, while the US method exhibited a flat response for the case with out-of-plane motion. In another comparative study using jets and wakes, Boomsma et al.[12] showed that indirect methods can yield a better uncertainty prediction with a better calibration using a distinct upper and lower bound for prediction. The analysis also revealed higher sensitivity for direct methods, although, it was shown that IM and CS methods can under-predict the standard uncertainty even when the systematic error is negligible.

Recently, Scharnowski et al. [13] proposed an uncertainty estimation method based on the loss-of-pairs due to out-of-plane motion. They quantified the loss-of-pairs as a ratio of the volume of the cross-correlation function to the volume of the autocorrelation function and proposed an uncertainty estimate based on the estimated loss-of-pairs. Optimizing this uncertainty prediction model for real experiments showed minimum error is achieved when loss of correlation due to out-of-plane motion is less than one.

In this work, we adopt an alternative approach and seek to quantify PIV measurement uncertainty directly from the information contained within the cross-correlation plane. The cross-correlation plane represents the distribution of probabilities of all possible particle image pattern displacements between consecutive frames, combined with the effect of the number of particles, mean particle diameter and effects that contribute to loss of correlation. In other words, the correlation plane is a surrogate of the combined effects of the various sources of error that govern the accurate estimation of a particle pattern displacement. The primary peak or the highest peak in the cross-correlation plane denotes the most probable displacement for a given particle image pattern. For an ideal shift between the particle patterns, a perfect cross-correlation peak can be represented by a convolution between a Dirac function (at the location of the shift) and the autocorrelation of particle image diameter. However any deviation in the peak shape is a manifestation of the errors influencing the measurement. Since, the standard uncertainty is typically defined as the standard deviation of all possible measurement values, we believe it is possible to directly estimate the uncertainty of each PIV measurement by the second order moment of the correlation plane. Hence, in this work we introduce a new method, the Moment of Correlation (MC), and establish the appropriate processing steps to extract the standard uncertainty from the cross-correlation plane. We demonstrate the sensitivity of the MC method to elemental error sources and compare its performance with existing methods (CS and IM) for synthetic and experimental data. This method has the benefit over those previously proposed in that limited additional pre- or post-processing is required, and it is not necessary to perform extensive processing-dependent calibration steps beforehand.

## 2  Methodology

The standard uncertainty is defined in section 2.1. We then derive the PDF of the displacement from the cross-correlation plane in section 2.2 and finally describe the methodology to extract the standard uncertainty from the PDF in section 2.3.

## 2.1 Definition of uncertainty

Uncertainty ($\pm u$) is the estimate of a range of values around the measurement that contain the true result, and bounds the true error. Usually, the uncertainty is provided at a defined "confidence interval", this means a certain percentage of data points will stay within the provided range. For example, the confidence interval within one standard deviation ($\sigma$) range for a Gaussian error distribution is 68% and within $\pm 2\sigma$ range is 95%. Standard uncertainty ($\sigma_u$) is defined as the one standard deviation ($\sigma$) level for the parent population of the variable [14], which is not required to be a Gaussian distribution. Therefore, the equation to calculate standard uncertainty can be written as follows (equation (15)):

$$\sigma_u^2 = E\left[(X-\mu)^2\right] = \int_x (x-\mu)^2 p(x) dx \qquad (1)$$

Where $\mu$ is the mean or expected value for $x$, and $p(x)$ is the probability distribution function (PDF).

## 2.2 Statistics of PIV correlation plane and uncertainty

Scharnowski et al. [15] showed that for an ensemble PIV correlation, the PDF of observed displacements in that ensemble, $p(d)$, can be calculated by deconvolving the contribution of the average particle image, $P_I$, from the ensemble averaged correlation ($<R>$) [15] (equation (15)):

$$\langle R \rangle = p(d) \otimes P_I \qquad (2)$$

We propose that for an instantaneous measurement, the PDF of possible displacement matches can be also computed by removing the particle image shape information ($P_I$). If image $a_2$ is obtained by shifting image $a_1$ by displacement $d$, as shown in equation (3), then using the Fourier shift theorem, the Fourier Transform (FT) ($\mathcal{F}$) of image $a_2$ can be written as shown in equation (4):

$$a_2(x) = a_1(x-d) \qquad (3)$$

$$A_2(r) = \mathcal{F}\{a_2\} = \mathcal{F}\{a_1(x-d)\} = A_1 \exp(-ird) \qquad (4)$$

In PIV, typically the displacement $d$ is estimated using Standard Cross Correlation (SCC) technique ($R = a_1 \otimes a_2, R = a_1 \otimes a_2$), which is evaluated in the Fourier domain using equation (5) as shown in Figure 1. Here, $R^*$ denotes the FT of the cross-correlation plane ($R$). The average particle image information $P_I$ can be estimated from the magnitude part of the cross correlation, in the frequency domain ($|R^*|$), as shown in equation (6), where $\overline{A_1}$ denotes the complex conjugate of the FT of the image $a_1 (A_1 = \mathcal{F}(a_1))$.

$$R^* = \mathcal{F}\{R\} = \overline{A_1(r)} \cdot A_2(r) = \overline{A_1} \cdot A_1 \exp(-ird) \qquad (5)$$

$$P_I \approx \mathcal{F}^{-1}\{|R^*|\} = \mathcal{F}^{-1}\{|\overline{A_1} \cdot A_2|\} \qquad (6)$$

So, $P_I$ can be removed by dividing $R^*$ by its magnitude ($|R^*|$) in the frequency domain and the Inverse Fourier Transform (IFT) of that ratio forms a Generalized Cross Correlation (GCC), plane, denoted by $G(x)$, as shown in equation (15).

$$G(x) = \mathcal{F}^{-1}\left(\frac{R^*}{|R^*|}\right) = \mathcal{F}^{-1}\left(\overline{A_1} \cdot A_2 / |A_1 \cdot A_2|\right) = \mathcal{F}^{-1}\left(\exp(-ird)\right) \qquad (7)$$

Since the FT is a linear operation, the remaining part is the summation of all possible matching shifts as described by equation (4), and therefore the GCC plane represents the PDF of candidate displacements. However, we consider the location of the primary peak (highest peak) as the most probable displacement, and given the displacement, the spread of the primary peak region is considered as the PDF of interest for our case. Therefore, the primary peak region in $G(x)$ (as shown in Figure 1) is the PDF ($p(x)$) of all possible matches in the correlated image pair that contribute to evaluation of the most likely displacement, multiplied by some constants having to do with the intensity level of the images correlated.

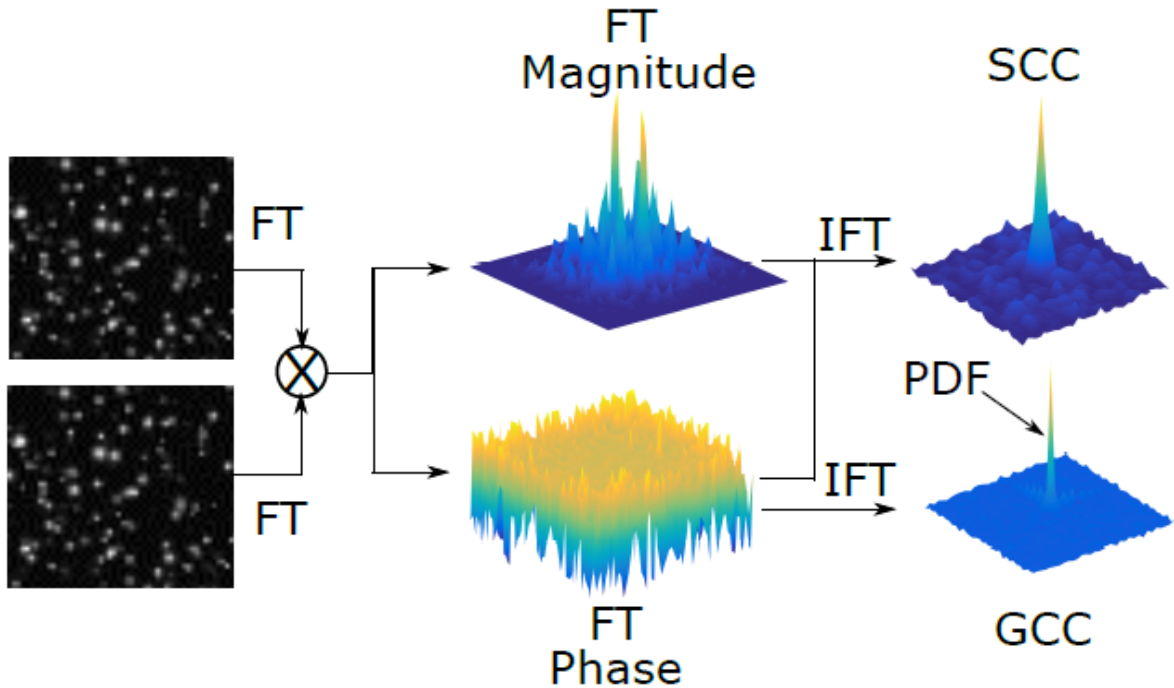

**Figure 1. Extracting PDF of displacement from PIV image pair cross correlation.**

Once the PDF of possible displacements is obtained, the second order moment about the primary peak, $X_p$, can be calculated as:

$$I_{xx}^2 = \int_X (x-X_p)^2 p(x)dx = \int_X (x-X_p)^2 G(x)dx \qquad (8)$$

Comparing equation (15) and (1), it is obvious that the standard uncertainty for a given PIV correlation can be expressed as $u = \sqrt{\int (x-X_p)^2 p(x)dx} = I_{XX}$. Therefore, the expected relationship between $I_{XX}$ and $u$ should be one-to-one.

However, calculating $I_{XX}$ directly is subject to large bias and random errors due to limited resolution in resolving the sharp primary peak in the normalized GCC plane. To compensate, we compute $I_{XX}$ by performing a Gaussian least square fit on the GCC plane convolved with a Gaussian function. The algorithm to find the standard uncertainty is described in the following section.

## 2.3 Moment of Correlation (MC) Algorithm

The Moment of Correlation algorithm, as described in Figure 2, extracts the standard PIV measurement uncertainty from the GCC plane. As a first step (Figure 2a) we convolve the GCC plane or the PDF with a 2d Gaussian function with zero mean and a selected standard deviation. We define the diameter of a Gaussian to be 4 times its standard deviation. The peak diameter estimated from the SCC plane is used as the convolving Gaussian diameter $(D_x, D_y)$. The convolved GCC plane $R_{conv}$ is given by

$$R_{conv} = G(x,y) \otimes \exp\left[-8\left(\frac{x^2}{D_x^2} + \frac{y^2}{D_y^2}\right)\right], \qquad (9)$$

where $D_x$ and $D_y$ are the Gaussian diameters in the $x$ and $y$ directions. For a large number of particles in an interrogation window the PDF ($G(x,y)$) can be reasonably approximated by a Gaussian distribution. Consequently $R_{conv}$ should also be a Gaussian.

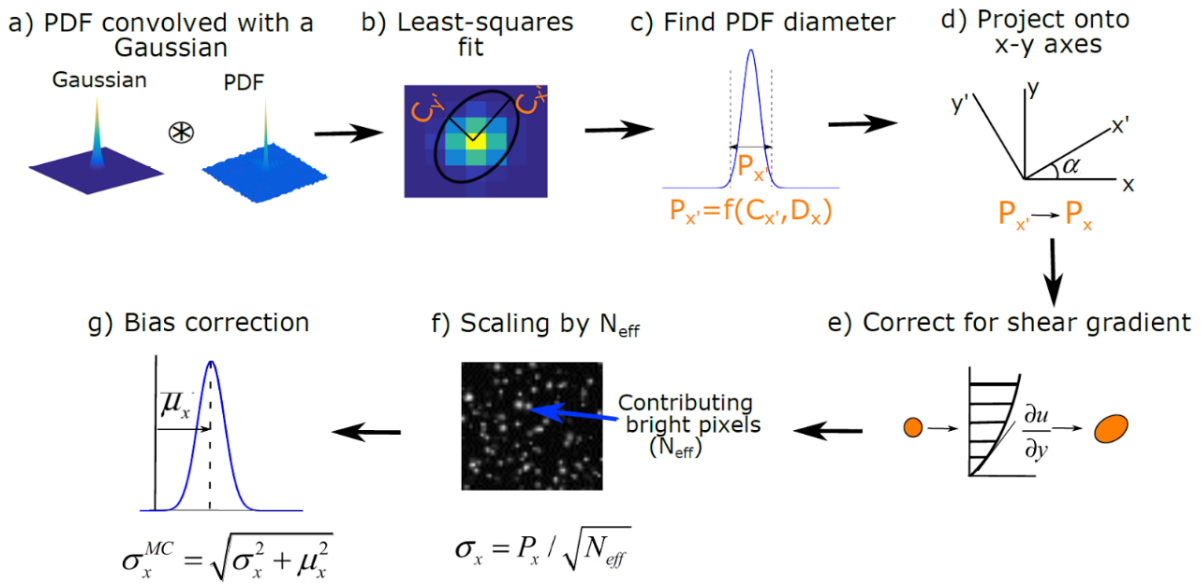

Figure 2: Algorithm to find standard uncertainty from PDF of displacements.

In the next step, Figure 2b, a Gaussian least squares fit is performed on the peak region of $R_{conv}$ to estimate the peak location $(X_c, Y_c)$ and its spread $(C_x, C_y)$. The general possibility of the $R_{conv}$ peak shape being elliptic Gaussian due to velocity gradients or the covariance of $\sigma_s$ and $\sigma_y$ is considered and thus the major axis $C_x$, minor axis $C_y$ and orientation α are estimated using the least squares fit (equation (15)).

$$\sum_{minimize} \left( R_{conv} - \frac{8}{\pi C_x C_y} \exp\left[ -8\left\{ \left(\frac{\cos\alpha(X-X_c) - \sin\alpha(Y-Y_c)}{C_x}\right)^2 + \left(\frac{\sin\alpha(X-X_c) + \cos\alpha(Y-Y_c)}{C_y}\right)^2 \right\} \right] \right)^2 \quad (10)$$

Once $C_x$ and $C_y$ are known, equation (15) is used to evaluate the PDF major axis $P_{x'}$ and minor axis $P_{y'}$ (Figure 2c):

$$\begin{aligned} P_{x'} &= \sqrt{C_x^2 - D_x^2} \\ P_{y'} &= \sqrt{C_y^2 - D_y^2} \end{aligned} \quad (11)$$

This relation (equation (15)) follows from the definition of convolution between two Gaussian functions. In step d (Figure 2d), the estimated $P_{x'}$ and $P_{y'}$ are projected from $x', y'$ on to $x$ and $y$ axis. The uncertainty or standard deviation $(P_x, P_y)$ is obtained by dividing the pdf diameter by 4 (equation (12)).

$$\begin{aligned} P_x &= \frac{1}{4}\sqrt{\cos^2\alpha P_{x'}^2 + \sin^2\alpha P_{y'}^2} \\ P_y &= \frac{1}{4}\sqrt{\sin^2\alpha P_{x'}^2 + \cos^2\alpha P_{y'}^2} \end{aligned} \quad (12)$$

The estimated standard deviation is corrected for velocity gradient using equation (13) as mentioned in Scharnowski et al.[15] (Figure 2e):

$$\begin{aligned} P_x &= \sqrt{P_x^2 - \frac{D_p^2}{16}\left(\frac{\partial u}{\partial y}\right)^2} \\ P_y &= \sqrt{P_y^2 - \frac{D_p^2}{16}\left(\frac{\partial v}{\partial x}\right)^2} \end{aligned} \quad (13)$$

Here $D_p$ represents the average particle diameter estimated from the cross-correlation plane. The standard uncertainty $(P_x, P_y)$ estimate thus obtained is much higher than the true uncertainty and requires a scaling factor. Since we are trying to estimate the uncertainty in the mean velocity (which we have assumed to be the peak of the PDF), not the uncertainty on a single particle match, it is expected that the uncertainty in the mean estimate should be reduced by the number of samples contributing to the mean. In this case the uncertainty is found to be appropriately scaled by the effective number of pixels ($N_{eff}$) correlating to produce the primary peak (Figure 2f). This factor is calculated by estimating the Mutual Information or MI [8] between the correlating windows. The MI is defined as the ratio of the cross correlation plane peak magnitude to the autocorrelation peak magnitude of one "average" particle and is equivalent to $N_I F_I F_O N_\Delta$ (product of $N_I$: number of particles in the window, $F_I$: fraction of particles lost due to in plane motion, $F_O$: loss of correlation due to out of plane motion and $N_\Delta$: loss of correlation due to local velocity gradients), which is just the number of particles contributing to the correlation. Thus, assuming a circular particle with area of $\frac{\pi}{4} D_p^2$, we define $N_{eff} = MI * \frac{\pi}{4} D_p^2$. Equation (14), then represents the standard uncertainty in x direction.

$$\sigma_x = \frac{P_x}{\sqrt{N_{eff}}} = \frac{P_x}{\sqrt{MI * \frac{\pi}{4} D_p^2}} \qquad (14)$$

In the last step (Figure 2g), we add a bias correction term to the random uncertainty estimate to get the overall standard uncertainty. In a multi-pass converged PIV processing, the shift between the two images should ideally be zero, which implies the estimated PDF distribution should have a peak at zero. Thus, non-zero peak location $X_c$, $Y_c$ is considered a bias in the uncertainty estimate. Hence, bias in $x$ direction is calculated as $\mu_x = X_c$ and the final MC method standard uncertainty, $\sigma_{MC}^x$ is given by equation (15).

$$\sigma_{MC}^x = \sqrt{\sigma_x^2 + \mu_x^2} \qquad (15)$$

Similarly, $\sigma_{MC}^y$ can be estimated.

## 3  Results

The methodology was first tested with synthetic images with varying magnitudes of several common error sources (section 3.1). The framework was also compared with IM and CS methods for more challenging flow cases in section 3.2. The details of the performance are given in the following sections.

## 3.1 Variation with elemental error sources

In order to evaluate sensitivity of the proposed algorithm to the primary PIV error sources, a set of artificial images were generated for a range of varying parameters. For the baseline conditions, images of 1024 by 1024 pixel size were generated with a seeding density of 0.05 particles per pixel (ppp) and particle image size of 2.6±0.13 pixels. The particle images were rendered within a 30 pixel wide uniform light-sheet, with 1% background noise, zero out-of-plane motion and uniform $x$ and $y$ displacements of 0.3 and 0.6 pixels respectively. For the individual cases, one parameter was varied at a time. The range of the parameters are as follows: displacement from 0 to 2 pixels in steps of 0.1 pixel, particle image diameter from 0.5 to 8 pixels in steps of 0.5 pixels, the $y$-shear rate was varied from 0 to 0.15 pixels/frame/pixel in steps of 0.025, background noise from 0.5 to 15% of maximum intensity with an increment of 1%, seeding density in the range of 0.005 to 0.15 ppp and the out-of-plane motion was varied from 0 to 40% of the light sheet thickness. For the shear case the image size was chosen as 256 by 4096 pixels to avoid large displacements at the edges in $y$-direction and also to have same total number of vectors as in other cases.

The images were processed using in-house open source code Prana [16], with two different window sizes of 64x64 and 128x128 pixels. In each case the windows were masked by a 50% Gaussian filter [17], such that the effective window resolution (WR) was 32 and 64 pixels respectively. For processing multi-pass iterative window deformation was used with a Standard Cross-Correlation (SCC). For each case the RMS error was compared to the RMS of the standard uncertainty estimate, obtained using the MC method. Each RMS value was calculated over 4096 samples.

Figure 3 shows the variation of the MC uncertainty estimate with primary PIV error sources. In each case the RMS error is denoted by the black line and the predicted uncertainty by the red line. The square and circular symbols denote the WR 32 and WR 64 cases respectively. For ideal prediction the RMS values of the error and predicted uncertainty should match perfectly. In this case, the MC method predicted uncertainty faithfully follows the RMS error trend and shows good sensitivity to the elemental PIV error sources. However, a bias of the order of 0.01 to 0.02 pixels is noticed in each case. Also, the degree of bias is relatively higher for the smaller window resolution case (WR32). This can be attributed to some bias in the estimate of the normalization factor $N_{eff}$ in the MC algorithm. Also, for bigger windows there are more effective correlating pixels, which statistically reduces the uncertainty on the sample mean. Overall, the response of the predicted uncertainty to the different error sources and its close agreement with the RMS error validates the MC method as a planar PIV uncertainty measurement tool.

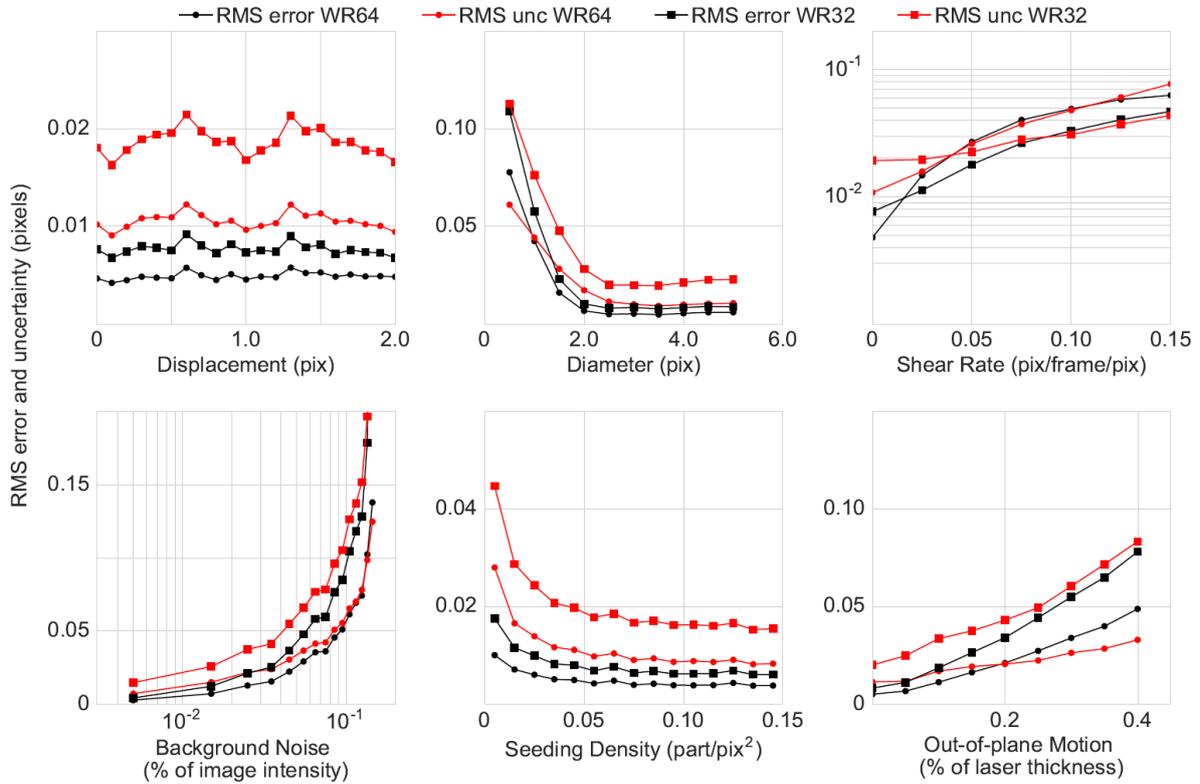

**Figure 3: Sensitivity of MC method to primary PIV error sources for two different window resolutions (32 and 64).**

## 3.2 Evaluation for complex flow fields (simulated and experimental test cases)

The MC framework was further tested for complex flow cases and the uncertainty estimates were compared with IM and CS predictions for each case. A total of five datasets were used. Two synthetic datasets, namely the Turbulent boundary layer (TBL) images from 2nd PIV Challenge (2003, Case B) [18] and the Laminar Separation Bubble (LSB) flow images with varying signal to noise ratio from 3rd PIV Challenge (2005, Case B) [19], were evaluated. Also, three experimental datasets of canonical flows were used for this analysis (cases C to E in Table 1). The details about the Stagnation Flow (SF) data can be found in Charonko et al. [5]. The Vortex Ring (VR) data is the central camera images of the case E in fourth PIV challenge [20]. Finally, the Jet Flow (JF) image set is taken from the same experiment as described in the "unsteady inviscid core" case of the collaborative uncertainty framework by

Sciacchitano et al. [11]. In each case the error analysis was done using a true solution, the details of which can be found in respective publications.

The images were processed using SCC with iterative window deformation for two different settings of window resolutions, WS1 and WS2, as described in Table 1. WS1 setting refers to a bigger final pass window resolution compared to WS2 setting. The PIV processing was done using Prana and DaVis 8.2, with MC and IM methods implemented in Prana and CS estimates obtained through DaVis. The number of passes and window overlap setting for each case are mentioned in Table 1. The final pass was processed without any validation or smoothing. Also, measurements with absolute error greater than 1 pixel were considered invalid and removed from any statistical analysis presented in the results. The following sections describe the overall error and uncertainty histogram, the agreement of the RMS error and uncertainty prediction, spatial variation of predicted uncertainties, and the uncertainty coverage obtained for each case in the test matrix using MC, IM and CS methods.

**Table 1: Description of test cases and processing parameters**

|  | **Case A** | **Case B** | **Case C** | **Case D** | **Case E** |
|---|---|---|---|---|---|
|  | Turbulent boundary layer (**TBL**) | Laminar separation bubble (**LSB**) | Stagnation flow (**SF**) | Vortex ring (**VR**) | Jet flow (**JF**) |
| **WS1** | 64x64 (75%-2) (87.5%-2) | 64x64 (75%-4) | 64x64 (75%-4) | 64x64 (75%-1 (87.5%-3) | 32x32 (87.5%-4) |
| **WS2** | 64x64 (87.5%-1) 32x32 (75%-3) | 64x64 (75%-1) 32x32 (50%-3) | 64x64 (75%-1) 32x32 (50%-3) | 64x64 (87.5%-1) 32x32 (75%-3) | 32x32 (75%-1) 16x16 (75%-3) |

## 3.2.1 Error and uncertainty histogram

Figure 4 shows the error and uncertainty histogram for each of the test cases. In Figure 4a, the error distribution is shown for cases A to E and for both WS1 and WS2 processing. The solid and the dashed black lines denote the errors obtained using Prana ($e_{Prana}$) and DaVis ($e_{DaVis}$) respectively. Figure 4b shows the uncertainty histogram for MC, IM and CS methods, overlaid on each other and are denoted by $\sigma_{MC}$, $\sigma_{IM}$ and $\sigma_{CS}$ respectively. Uncertainty distributions are plotted only on the positive x-axis, assuming a symmetric uncertainty curve bounds the error histogram on the negative x-axis. This assumption is correct for these cases, as the error histogram is symmetric about zero with a maximum bias of -0.015 pixels observed for case D. The uncertainty histogram for MC, IM and CS methods are plotted together with red, cyan and violet colors respectively. The CS uncertainty distribution for the laminar separation bubble (case B) shows multiple peaks, which may be an effect of the decreasing signal to noise ratio in those images. However, the error distribution does not show multiple peaks in its distribution and likewise the MC uncertainty prediction also shows a single peak in the histogram. For the stagnation flow case (case C), all three methods show two distinct peaks in their distribution, owing to the different x and y uncertainty values plotted together. In all the cases, x and y error and uncertainty distributions are lumped into a single error and uncertainty vector, which are then divided into 40 and 60 bins respectively, to draw the histogram.

The vertical lines show the RMS values of the error and uncertainty distributions. The basis of comparison is that, for an ideal prediction, the RMS of error and uncertainty distributions should match each other [11]. Thus, the RMS error lines in Figure 4a are repeated in Figure 4b for ease of comparison. The vertical RMS error lines for Prana and DaVis match each other nearly perfectly with a maximum difference of less than 0.01 pixels. The violet dashed line (RMS of $\sigma_{CS}$) should be compared to the black dashed line (RMS of $e_{DaVis}$), while the solid red (MC) and the cyan (IM) vertical lines should be compared to the Prana RMS error (solid black). However, since the RMS errors are almost identical, the RMS uncertainties can be compared with respect to each other as well. For ease of comparison, Table 2 lists the RMS values of error and uncertainty distributions for all test cases. The predicted uncertainties reasonably match the RMS error with a maximum deviation of about 0.02 to 0.03 pixels for the experimental cases (SF, VR and JF). For the synthetic cases (TBL and LSB) the RMS uncertainties are within ±0.01 pixels of the RMS error values. The RMS of $e_{Prana}$ and $e_{DaVis}$ are about 0.03 pixels for cases A and B, while it is higher (0.05 to 0.07 pixels) for the experimental cases (case C, D and E). For the vortex ring (case D), all the methods under predict the standard uncertainty while in other four cases, depending on the processing, the RMS uncertainties are seen to both underestimate and overestimate the RMS error (Table 2). Since the reference solution for the VR case was obtained using a multi-camera tomographic reconstruction, the planar (front on) camera image processing may incur some bias error with respect to the "true" solution. Such a systematic error can influence the consistent underprediction of the estimated uncertainties.

The methods show differences in their predictions for the WS1 and WS2 settings (Figure 4). With the WS1 setting, the RMS error and the RMS uncertainty predicted using MC method closely match each other for cases A, B, and C, but the RMS estimation under predicts the true errors for cases D and E. In contrast, for this window resolution the CS and IM methods under predict the RMS error in cases A and D and match the MC estimates in other cases. For the WS2 setting, however, MC method over predicts the RMS error for cases A, B, and C but matches the RMS error closely for cases D and E. In this window resolution, the CS and the IM estimates match the RMS error closely for case A and E. Overall, all three methods closely predict the correct standard uncertainty with MC method doing a better job in cases A, C, and D for the WS1 setting and cases D and E with the WS2 processing.

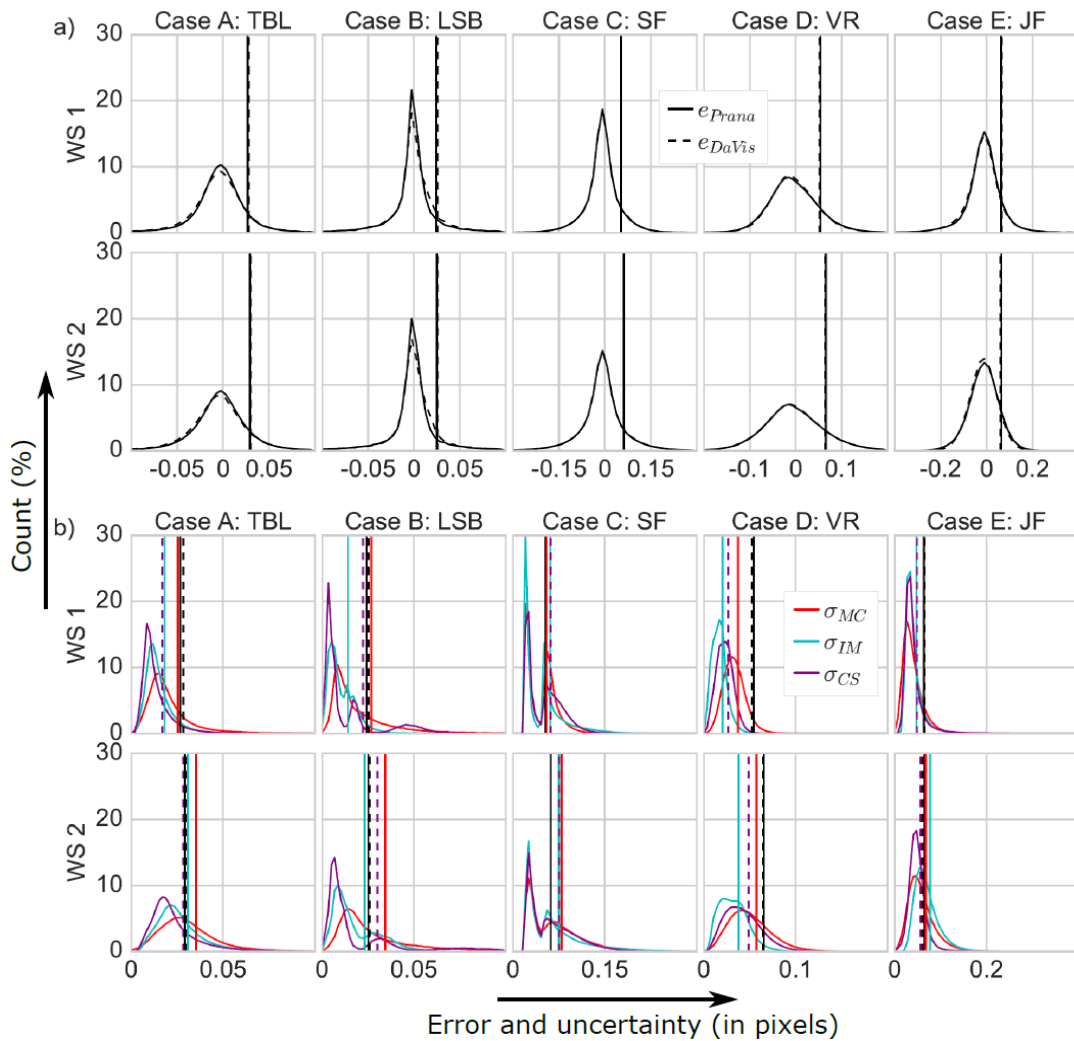

Figure 4: Error and uncertainty histogram comparing MC, PD and CS performance for the five test cases.

**Table 2: Comparing the RMS error and uncertainty values across different methods for the five test cases.**

|     | RMS (Pixels) | Case A: TBL | Case B: LSB | Case C: SF | Case D: VR | Case E: JF |
|---|---|---|---|---|---|---|
| WS1 | $e_{Prana}$ | 0.027 | 0.024 | 0.053 | 0.054 | 0.064 |
|     | $\sigma_{MC}$ | 0.025 | 0.027 | 0.055 | 0.037 | 0.047 |
|     | $\sigma_{IM}$ | 0.018 | 0.014 | 0.061 | 0.020 | 0.047 |
|     | $e_{DaVis}$ | 0.028 | 0.025 | 0.053 | 0.053 | 0.065 |
|     | $\sigma_{CS}$ | 0.017 | 0.022 | 0.061 | 0.026 | 0.048 |
| WS2 | $e_{Prana}$ | 0.029 | 0.025 | 0.062 | 0.065 | 0.064 |
|     | $\sigma_{MC}$ | 0.035 | 0.034 | 0.079 | 0.057 | 0.067 |
|     | $\sigma_{IM}$ | 0.031 | 0.023 | 0.075 | 0.038 | 0.077 |
|     | $e_{DaVis}$ | 0.030 | 0.026 | 0.062 | 0.065 | 0.060 |
|     | $\sigma_{CS}$ | 0.028 | 0.030 | 0.075 | 0.049 | 0.056 |

### 3.2.2 Predicted vs expected uncertainty

The predicted uncertainties have a distribution and not a single value due to an inherent uncertainty in the PIV uncertainty estimation. This is attributed to the degree of overlap between correlating particles [10]. Thus, to analyze the distribution of uncertainty the uncertainty values are divided into 8 bins, and for measurements falling in each bin the RMS error and uncertainty values are plotted. Figure 5 shows a direct comparison between the RMS error or the expected uncertainty versus the predicted uncertainty for each method. For an ideal prediction the graph should be a line with slope equal to 1. The deviation from black dashed line in the plots is indicative of the amount of failure in each prediction. Hence, when the predicted uncertainties lie below and to the right of the 1:1 reference line, the predicted uncertainty estimate is under-predicting the true error distribution, and when it is above and to the left the true errors are smaller than the prediction. For the WS1 processing, the MC method closely follows the dashed line, especially for cases A to C. However, the deviation increases for higher uncertainty bins. For the first two cases the $\sigma_{IM}$ and $\sigma_{CS}$ under predict the 1:1 line for the lower uncertainty values but do a better job for WS2 processing. For case C, the $\sigma_{MC}$ outperforms the other methods, while in case D, all the methods under predict the true error distributions line. Thus, Figure 5 shows that the predicted uncertainty distributions match the

RMS error closely over the whole range, except for some deviation in the higher uncertainty bins. Comparing these results to Figure 4a, it can also be seen that the regions of greatest deviation between the predicted uncertainty and true errors lie in the tails of the error distribution where there are much fewer samples to draw from, and thus the statistical estimates are more sensitive to outliers.

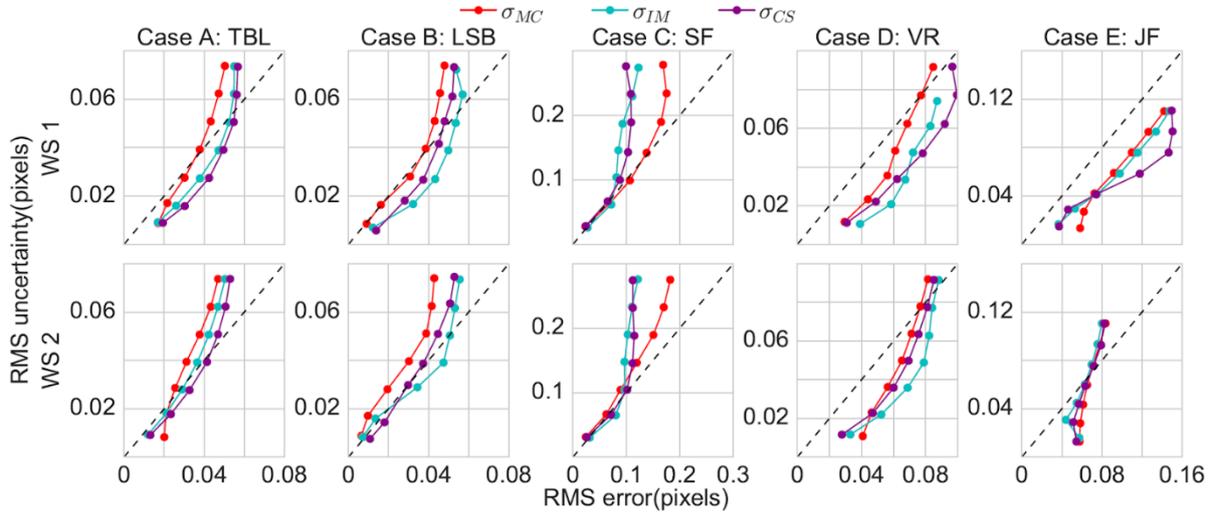

Figure 5: Comparing RMS error versus RMS of the predicted uncertainty for each method (MC, PD and CS) for the five test cases.

### 3.2.3 Spatial variation in RMS error and uncertainty

The spatial variation in RMS error and RMS uncertainty for a specific location in the flow field is analyzed. A particular *x* or *y* grid coordinate value is used to select a vertical or horizontal line cutting across the flow field and then the RMS of the error and uncertainty values (for x-component of velocity) across the time series is plotted along that line as shown in Figure 6. In each RMS value calculation, any measurement with error greater than 1 pixel is considered as invalid and not taken into account. For cases A to C, a vertical line (*x* value set equal to the mid-point) along the middle of the grid of vectors is selected. Specifically for case B, the lowest SNR case is not included as it increased the noise level in the RMS spatial profiles without adding any significant trend in the comparison.

For case D, a horizontal line through the top vortex core is selected. For the jet flow case, a vertical line cutting across the horizontal jet and towards the right hand edge (downstream of the jet at x=380 pixel of the true solution grid) is selected. For case A, higher error and uncertainty values are noted near the wall (normalized coordinate 0). In this case the MC method is seen to be more sensitive to the spatial variation in the RMS error. For the laminar separation bubble case (case B), the RMS is taken across the decreasing SNR cases. The MC method shows better sensitivity to the spatial variation but over predicts the standard uncertainty for the WS2 processing (smaller windows). For the stagnation flow (case C), due to 3D flow and high shear rates near the wall, at near wall normalized coordinate 1 the error and uncertainty values reach about 0.1 pixels. The vortex ring case (case D), shows high fluctuations in error values near the vortex core at location 0.5. For both cases C and D, all three methods show reasonable variation corresponding to the error curves, however, the MC method clearly shows higher sensitivity to the error peaks. Finally, for the case E, the uncertainty curves show poor response for all methods in the shear layer region for WS1 processing with larger final interrogation windows. For the smaller window resolution (WS2) the IM and the MC method matches the RMS error curve and shows good sensitivity to variation in spatial error, but CS method shows lesser response to the error peaks in the shear layer. Overall, the MC method shows better sensitivity to the spatial variation in the RMS error for all the cases.

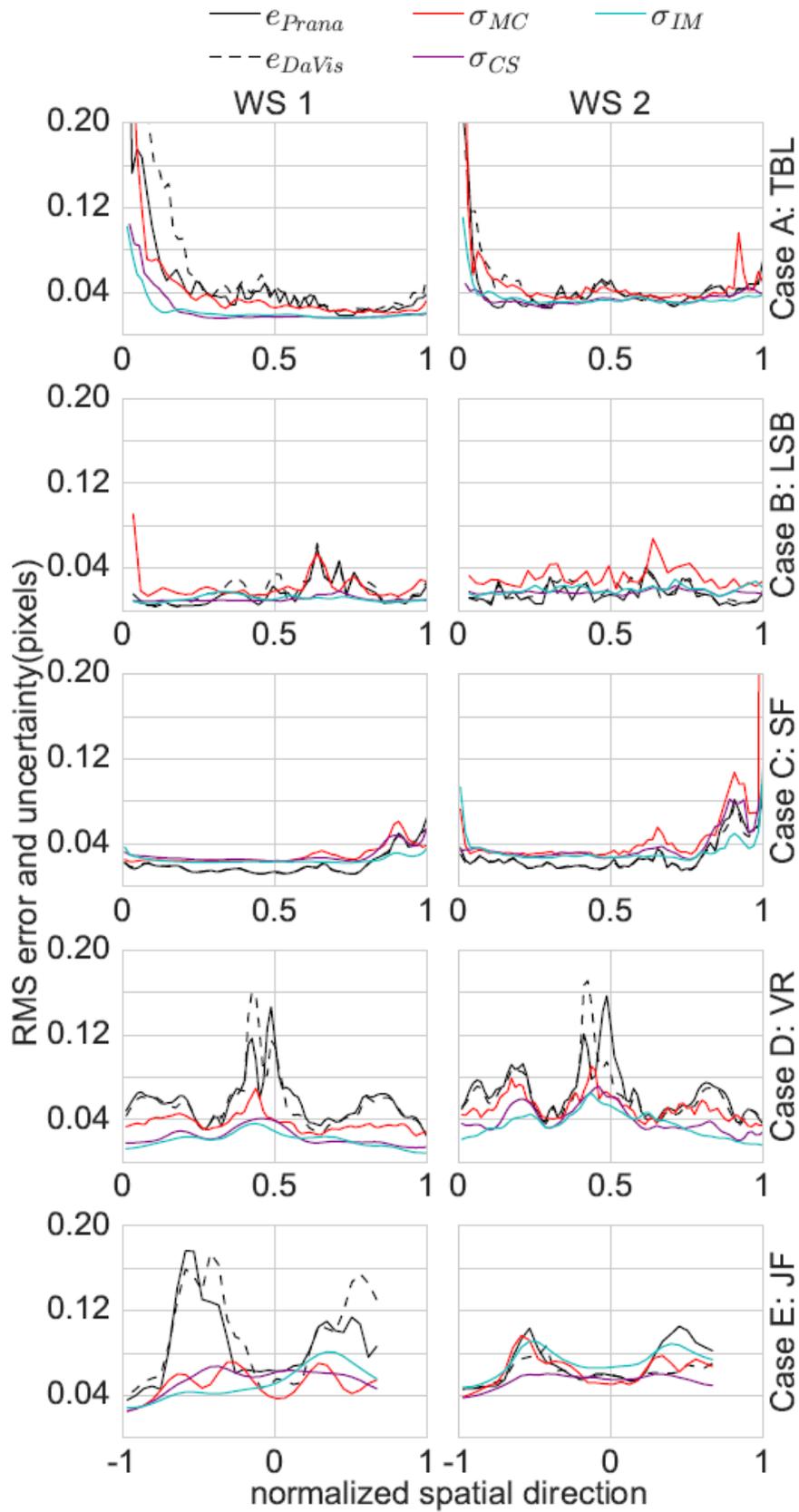

**Figure 6:** Comparing the spatial variation in RMS error and the RMS of the estimated uncertainties using MC, IM and CS methods for the different test cases.

## 3.2.4 Uncertainty coverage

Another measure of successful uncertainty prediction is the uncertainty coverage, which denotes the percentage of measurements for which the error lies within the uncertainty bound. For a Gaussian error distribution, this should be ideally 68.5%. However, the error distribution can deviate from a Gaussian distribution and since coverage by definition is the fraction of measurement errors falling within ±σ (standard deviation) of the error distribution, such a measure is independent of any specific type of distribution for the error. The target coverages are thus calculated from the true error distributions without the assumption of normality and are shown as small black squares in Figure 7. Figure 7a) demonstrates the coverage values for each method and for all the different flow cases separately, while Figure 7b) shows the expected and predicted coverage bars combined across all the different flow cases, for WS1 and WS2 processing. The expected or target coverage for all cases is between 69% and 81%, with the VR case expected coverage (square markers) being closest to the 68.5% mark. Expected values higher than 68.5% indicate that the true error distributions are less compact than Gaussian, and have longer tails. The WS2 processing is denoted by the hatched bars for each method. In general, the WS1 processing shows a lower coverage for all cases except for the case C, meaning the uncertainty is being underestimated. For the vortex ring case (case D), all the methods show a reduced coverage of about 26% to 55%, with MC performing better compared to IM and CS. This could indicate a failure of the uncertainty estimate or suggest a systematic bias in the reference solution which was derived from an auxiliary tomographic PIV measurement. For the cases A and B, the IM and CS methods under predict the coverage for WS1 processing, whereas MC method predicts a coverage of around 62% and 78%, respectively, the latter almost matching the true target coverage of 81%. For TBL and LSB cases with WS2 processing, MC method perfectly matches the expected coverage of 75% for the first case and over-predicts the expected coverage of 81% by about 9% for the second case. In contrast, IM and CS methods yield a coverage of 68% and 60% for case A (expected coverage 73%) and 73% and 63% for case B (expected coverage 81%), with WS2 processing. For the jet case (case E) the, the predictions for WS2 show better coverage compared to the WS1 processing, with IM method predicting the closest coverage match (71%). For the stagnation flow case, all the methods successfully predict a coverage of about 72% to 77%, which is at worst within 4% of the expected coverage (76%).

Figure 7b) compares the uncertainty coverage over all the measurement points, irrespective of any particular flow characteristics, emphasizing on the statistical performance.

The plot clearly brings out that for each window size processing, the MC method predicts the target coverage the closest as well as the fact that, WS2 processing in general yielded higher coverage compared to WS1 processing. It should be noted that the coverage does not capture the local variation in uncertainty prediction that was discussed earlier, however in an overall statistical sense better coverage usually indicates a better prediction and is a useful benchmark.

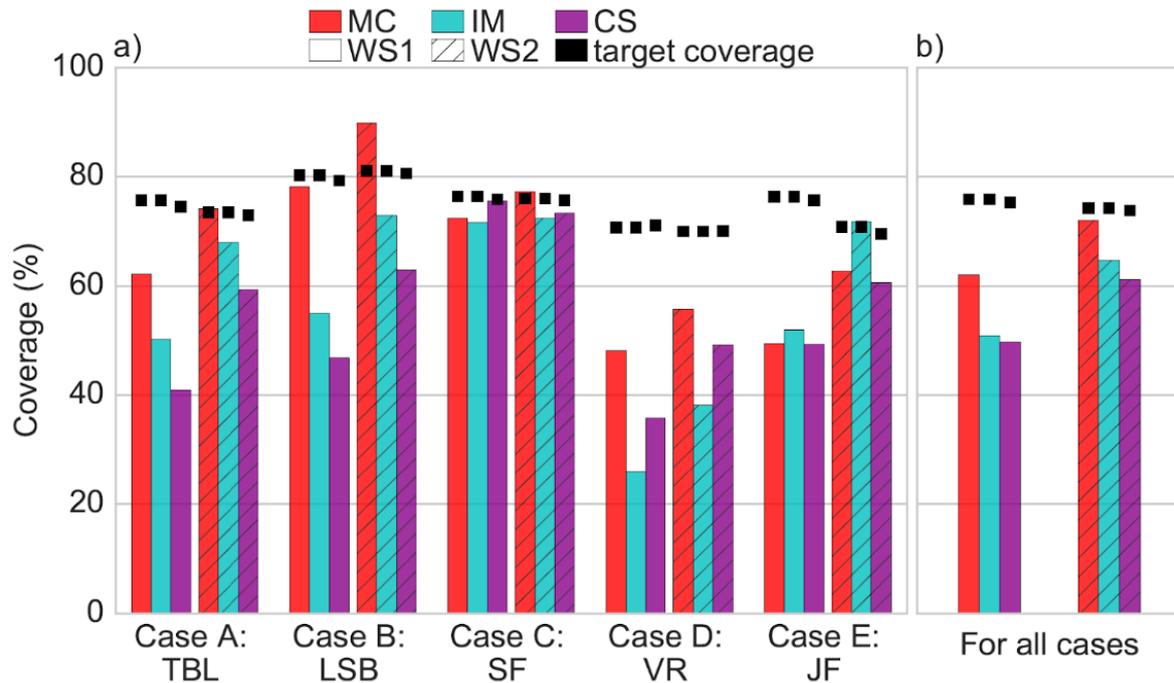

Figure 7: Grouped bar chart for standard uncertainty coverage using MC, IM and CS methods for WS1 and WS2 processing. a) for different flow test cases, b) for all cases combined. The target coverages calculated from the true error distributions are shown as black squares for each case.

## 4 Conclusion

A framework to extract the PIV uncertainty directly from the cross-correlation plane is provided herein. The PDF of all possible displacements that influence the final velocity prediction is first extracted from the instantaneous PIV correlation plane; this PDF is then convolved with a suitable Gaussian to reliably estimate the PDF diameter. The standard uncertainty is then determined using a least-squared Gaussian fit on the primary peak region of the convolved Gaussian plane accounting for any peak stretching or rotation. The final estimate is normalized by the effective number of pixels contributing to the cross-correlation peak. The present method shows strong agreement with the RMS error trends for each primary PIV error source. Further analysis with more complex flows revealed good agreement with the expected uncertainty distributions. The proposed method predicted the RMS error better than the existing

IM and CS methods, especially for the processing with larger window sizes. However, for lower window sizes the method over-predicted the standard uncertainty for the first two cases compared to the IM and CS estimates. The MC method showed better sensitivity to spatial variation in error compared to IM and CS methods for all cases. The standard uncertainty coverage predicted by the MC method was higher than the IM and CS method coverage, for most of the cases. A bias error of about 0.02 pixels was noticed for the MC method in the simulated cases. This bias error may be related to any bias in the estimated number of correlating pixels or in difficulty in sizing extremely small PDF peaks. Overall, after analyzing a wide range of test cases and the sensitivity of the predicted uncertainty to the variation in error sources, the MC method establishes itself as successful planar PIV uncertainty prediction tool and provides a framework to estimate cross-correlation uncertainty even in 3D cross-correlation.